\documentclass[prl,twocolumn,aps,floatfix,superscriptaddress,10pt]{revtex4-2}
\usepackage{amsmath}
\usepackage{amssymb}
\usepackage{graphicx}
\usepackage{color,verbatim,comment}
\usepackage{xcolor}
\usepackage{soul}
\usepackage{xr}
\usepackage{hyperref}
\usepackage{xr-hyper}
\usepackage{ulem}

\newcommand{\ba}{\begin{array}}
\newcommand{\ea}{\end{array}}

\begin{document}

\title{
Pseudovorticity of 2+1D optical solitons 
}

\author{Ludovica Falsi}
\affiliation{Dipartimento di Fisica, Universit\`a di Roma “La Sapienza”, 00185 Rome, Italy}
\affiliation{Enrico Fermi Research Center (CREF), Via Panisperna 89A, 00184 Rome, Italy}
\author{Giuseppe Agostino}
\affiliation{Dipartimento di Fisica, Universit\`a di Roma “La Sapienza”, 00185 Rome, Italy}
\author{Alberto Villois}
\email{a.villois@uea.ac.uk}
\affiliation{School of Engineering, Mathematics and Physics, University of East Anglia, Norwich Research Park, Norwich, NR4 7TJ, United Kingdom}
\author{Francesco Coppini}
\affiliation{Dipartimento di Fisica, Universit\`a di Roma “La Sapienza”, 00185 Rome, Italy}
\affiliation{Department of Mathematics, State University of New York at Buffalo, Buffalo, New York 14260-2900}
\author{Paolo M. Santini}
\affiliation{Dipartimento di Fisica, Universit\`a di Roma “La Sapienza”, 00185 Rome, Italy}
\affiliation{Istituto Nazionale di Fisica Nucleare (INFN), Sezione di Roma, Piazz.le Aldo Moro 2, I-00185 Roma, Italy}
\author{Miguel Onorato}
\affiliation{Dipartimento di Fisica, Universit\`a degli Studi di Torino, 10125 Torino, Italy}
\affiliation{Istituto Nazionale di Fisica Nucleare (INFN), Sezione di Torino, 10125 Torino, Italy}
\author{Aharon J. Agranat}
\affiliation{The Institute of Applied Physics, The Hebrew University, Jerusalem 91904, Israel}
\author{Stefano Trillo} 
\affiliation{Department of Engineering, University of Ferrara, 44122 Ferrara, Italy}
\author{Eugenio DelRe}
\affiliation{Dipartimento di Fisica, Universit\`a di Roma “La Sapienza”, 00185 Rome, Italy}

\begin{abstract}
In the hydrodynamic representation  of a quantum fluid or optical field, vorticity vanishes wherever the phase is well defined, and is instead localized at phase singularities, or quantum vortices. Pseudovorticity, by contrast, characterizes local rotational structures, even in regions without  singularities or net orbital angular momentum. 
We study both experimentally and numerically pseudovorticity in photorefractive solitons and show that  a detailed phase and amplitude analysis unveils a complex rotational flow dynamic: bright 2+1D solitons are found to carry a pseudovorticity dipole, while quadrupoles emerge in soliton fusion.  The phenomenon, also explained using geometrical considerations, suggests a general picture according to which stable high-dimensional solitons naturally carry a hierarchy of pseudovorticity multipoles, encoded in the local perturbed phase and amplitude.  

\end{abstract}

\maketitle

\noindent {\it Introduction. } Optical waves can induce rotation of absorbing or reflecting particles~\cite{Bruce2021}. This effect is generally attributed to the transfer of angular momentum, either spin or orbital angular momentum (OAM)~\cite{Allen2016, He1995, Friese1998, Forbes2024, Neshev2004}. However, optical torque can also arise in the absence of OAM, even for fields with zero net angular momentum, as first pointed out by Berry~\cite{Berry2009}. Such effects were first studied in superpositions of linear wavefields, which can sustain finite \textit{current vorticity} without phase singularities, and hence without topological charge. More generally, optical torque originates from the local spatial structure of the field: when phase and amplitude gradients are not collinear, the resulting momentum flux becomes spatially inhomogeneous, producing a net torque on absorbing particles.

This perspective can be formalized in terms of the momentum density of light. In optical trapping, forces are typically decomposed into conservative intensity-gradient contributions and a nonconservative radiation-pressure component~\cite{Ashkin1992}. The momentum density depends not only on intensity but also on the spatial structure of the phase, giving rise to transverse forces associated with phase gradients, even in the absence of intensity variations~\cite{Roichman2008}. This description extends to vectorial light fields, where spatial variations of polarization introduce additional terms to the momentum density, including spin-dependent and spin-curl components inducing forces and torques in the absence of orbital angular momentum~\cite{Ruffner2012}. More generally, optical forces and torques are governed by the local momentum and spin densities of the field~\cite{Toftul2026}, and are therefore determined by its full spatial structure and are intrinsically nonconservative.

A natural quantity capturing this mechanism is pseudovorticity, defined as the current vorticity of a complex field. It has been introduced as a diagnostic tool to identify vortex structures in quantum fluids described by a complex order parameter~\cite{Villois2016, Rorai2016}. Light propagating in nonlinear media exhibits analogous hydrodynamic behavior, including collective excitations and effective interactions~\cite{Glorieux2025, Dieli2026, Vocke2015}. While these connections have been extensively explored in the presence of phase singularities, the role of pseudovorticity in vortex-free nonlinear wavefields remains largely unexplored. In this regime, nontrivial current vorticity does not arise from topological defects, but instead reflects the intrinsic phase–amplitude structure of the field, even in the absence of singularities.

In contrast to linear waves, where such phase–amplitude mismatch must be externally imposed, it is expected to arise naturally and play a dominant role in nonlinear waves, such as optical solitons, where phase and amplitude are intrinsically coupled around a stable localized state. Solitons are self-trapped wave packets in which diffraction or dispersion is balanced by nonlinear self-focusing or self-phase modulation. While low-dimensional solitons, such as temporal solitons in optical fibers~\cite{Hasegawa1989, Akhmediev1997, Kibler2010, Dudley2006, Dudley2014}, cannot support pseudovorticity, it can arise in higher-dimensional 2+1D solitons, such as those observed in photorefractive crystals~\cite{Chen2012, DelRe1998, DelRe2009}. In this context, pseudovorticity should develop as soliton-supporting conditions are approached, reflecting the phase–amplitude dynamics associated with the self-trapped state.

Here, we present direct experimental and numerical evidence of pseudovorticity in bright 2+1D optical solitons and their interactions. We show that these systems exhibit a hierarchy of pseudovorticity multipoles, with dipolar structures arising in self-trapped solitons and quadrupolar patterns emerging both as the input beam deviates from the stationary state and during soliton–soliton interactions. This hierarchy reflects the underlying phase–amplitude structure of the field, with different multipolar components associated with distinct local perturbations of the soliton solution. These structures arise from the interplay between the nonlinear response and perturbations of the soliton profile, including anisotropic and nonlocal corrections. Pseudovorticity thus provides a unified framework to characterize the phase–amplitude coupling in high-dimensional solitons, revealing a hidden dynamical structure beyond conventional intensity or phase descriptions.

\noindent {\it Pseudovorticity of 2+1D Solitons.} 
In general, starting from a complex field written in terms of Madelung transform $\psi = \sqrt{\rho} e^{i \phi}$ that can describe the order parameter in a quantum fluid, a mean-field wavefunction, or an optical field, the pseudovorticity $\boldsymbol{\omega}_{\mathrm{ps}}$ associated with $\psi$ is defined as the curl of the current $\mathbf{j} \equiv \rho \mathbf{v}$, where $\mathbf{v}=\nabla \phi$ is a velocity field \cite{Berry2009,Villois2016,Rorai2016}:
\begin{equation} \label{PV1}
   \boldsymbol{\omega}_{\mathrm{ps}} = \nabla \times \mathbf{j}= \nabla \times (\rho \mathbf{v}),
\end{equation}
which yields the equivalent expression 
\begin{equation} \label{PV2}
   \boldsymbol{\omega}_{\mathrm{ps}} =  \nabla \rho \times \nabla \phi.
\end{equation}
To date, pseudovorticity has been proven to be a pivotal tool to track quantum vortices in 3D space in superfluids with dynamics ruled by Gross-Pitaevskii equation with repulsive (i.e., \textit{defocusing}) interactions, allowing for investigating their reconnection \cite{Villois2020,Serafini2017} and characterizing quantum turbulent regimes.
In this letter, we address a completely different problem, namely, whether pseudovorticity can emerge, being detectable in state-of-the-art experiments, from conventional (ground state or Townes) optical bright solitons of \textit{focusing} media, described by the following 2+1D generalized dimensionless nonlinear Schr\"odinger equation
\begin{equation} \label{NLSE}
i \, \partial_z \psi + \nabla_{\perp}^2 \psi + F(|\psi|^2)\psi=0.
\end{equation}
In this case the light field is equivalent to a compressible, inviscid, irrotational fluid with density $\rho=|\psi|^2$ and velocity $\mathbf{v}=\nabla Arg(\psi)$, subject to a (negative \cite{Gurevich1970}) effective pressure arising from the nonlinear term $F(\rho)$. The definition in Eq. (\ref{PV1}) and Eq. (\ref{PV2}) then holds with $\nabla \rightarrow \nabla_{\perp}=\partial^2_x+\partial^2_y$, and the pseudovorticity has a single component along $z$, i.e., it becomes effectively the scalar $\omega_{\mathrm{ps}} =  \nabla_{\perp} \rho(x,y) \times \nabla_{\perp} \phi(x,y)$.

\begin{figure}
    \centering
    \includegraphics[width=1\linewidth]{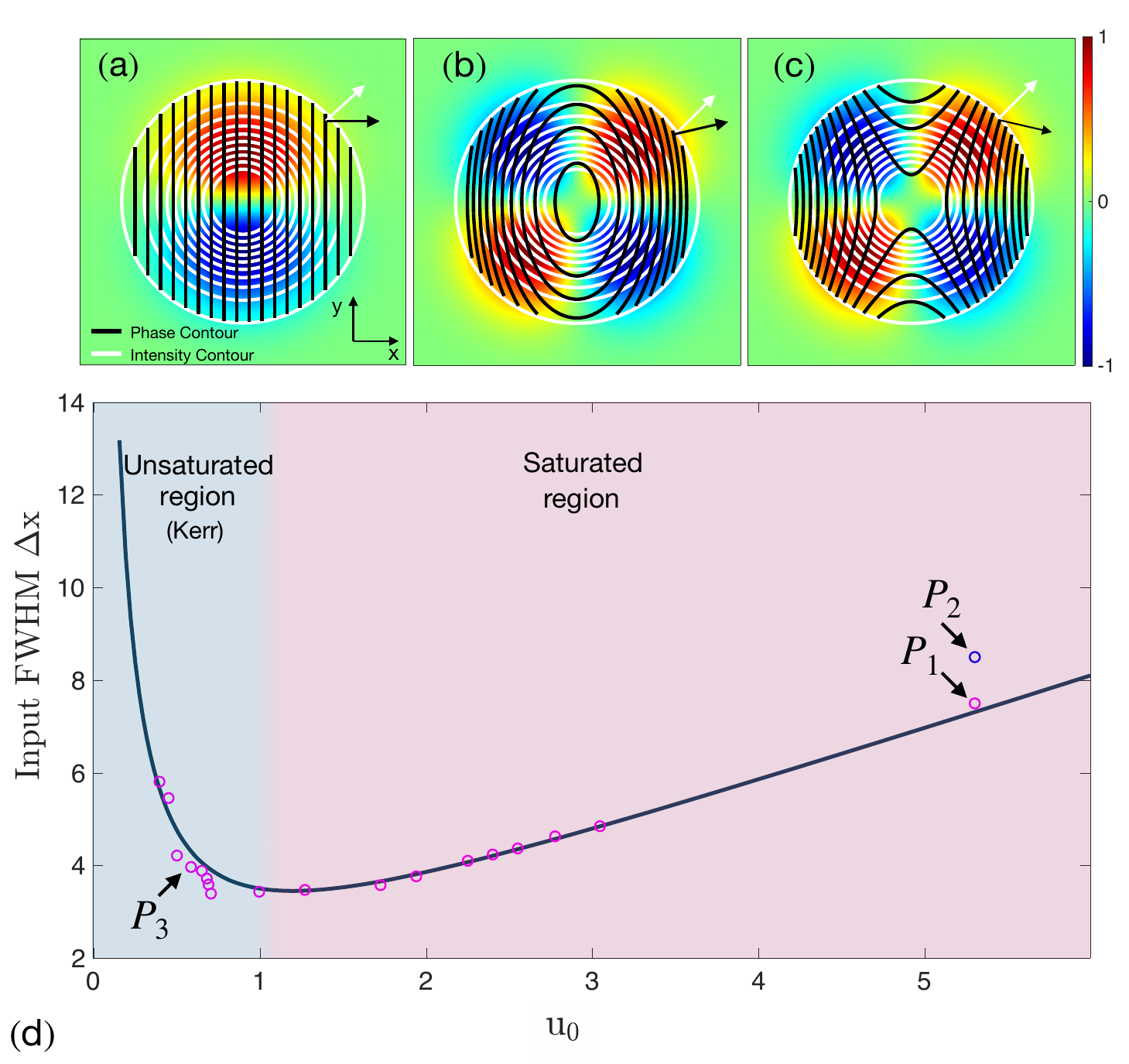}
    \caption{Pseudovorticity and 2+1D soliton existence conditions. Pseudovorticity (a) dipole for a beam with a linear phase gradient $(\phi = vx)$, (b,c) quadrupole for an elliptical (b) or hyperbolic (c) phase perturbation.
(d) Measured normalized soliton FWHM $\Delta x$ vs. amplitude $u_0$ (open circles) superimposed on the theoretical existence curve from Eq. (3) (solid line); $P_1$ and $P_3$ stand for soliton-supporting launching conditions in Kerr-like ($u_0 \lesssim 1$) and saturated ($u_0 \gtrsim 1$) regimes, respectively, while $P_2$ stands for overfocusing, not leading to a soliton.}
    \label{fig:enter-label}
\end{figure}
We begin by considering a stable soliton of Eq. \eqref{NLSE} with transverse dependence (see also end matter) $\psi(x,y) = \sqrt{\rho_s(r)}~e^{i \phi(x,y)}$, $r \equiv \sqrt{x^2+y^2}$, where $\rho_s(r)$ is a radially symmetric density profile and $\phi(x,y)=\phi_0=0$ is a flat phase (set to zero, without loss of generality, due to invariance under arbitrary phase rotation). We next consider lowest order perturbations of the phase field 
\begin{equation}
\phi(x,y)\approx
\sum_{i=1}^{2} b_i x_i
+\frac{1}{2}\sum_{i,j=1}^{2} a_{ij}\,x_i x_j
\label{perturbation}
\end{equation}
where \mbox{$\mathbf{x}=(x_1,x_2)=(x,y),\; b_i=\partial_i\phi|_{\mathbf{x}=0},\; a_{ij}=\partial_i\partial_j\phi|_{\mathbf{x}=0}$}.

In Fig.1a, we illustrate the case in which a localized field has a velocity along a (horizontal) $x$ direction, i.e., \(\phi(x) = v x\) ($b_1=v$, $b_2=0$, $a_{ij}=0$).  The concentric intensity contours (constant $\rho$, white lines) intersect the linear phase contours (constant $\phi$, black lines) at different angles.  The result is maximal when the two gradients $\nabla \phi$ and $\nabla \rho$ are orthogonal, i.e., along the $y$-axis, and vanishes along the $x$-axis where the gradients are collinear. Consequently, the pseudovorticity points along the $z$-axis, being positive above the soliton center, negative below, and zero along the direction of motion, thereby forming a dipole oriented orthogonally to the velocity direction (see also end-matter for a similar alternative argument).

The second-order term in \eqref{perturbation} corresponds to conic section phase contours. Since, for simplicity, we are analyzing the case of perfectly radial amplitude, we can, without loss of generality, study the case in which the mixed term $a_{xy}$ is zero, since it can be absorbed by a rotation of the plane. We thus have two cases: either $a_{xx}$ and $a_{yy}$ have the same sign or they have opposite signs. In Fig.1b we illustrate the first case (elliptical), where \( (a_{xx}, a_{yy}) > 0\). The phase profile 
$\phi(x,y)\approx a_{xx}x^2  + a_{yy}\,y^2$ contour lines are ellipses centered at the origin. When superimposed on the radially symmetric intensity field, the misalignment between $\nabla_{\perp} \phi$ and \(\nabla_{\perp} \rho\) leads to a pseudovorticity structure with four lobes aligned along the coordinate axes. The angular dependence results in a quadrupolar pattern with alternating sign in each quadrant, and nodal lines along the coordinate axes. The second case, in which the signs of $a_{xx}$ and $a_{yy}$ are opposite, leads to hyperbolic contour lines is illustrated in Fig.1c, for \(a_{xx} >0, a_{yy} < 0\),  \(\phi(x,y)\approx |a_{xx}|x^2  - |a_{yy}|\,y^2\). 
The resulting pseudovorticity structure is similar to that of the previous case, as illustrated in Figures Fig.1b and Fig.1c (see details in Interpretation of the multipolar structure in End Matter) 
  In both cases, the origin remains a stationary point of the phase, and the quadrupolar structure emerges from the angular curvature of the phase fronts. In general, second-order phase perturbations naturally give rise to pseudovorticity patterns with distinct quadrupolar symmetry, determined by the local geometry of the phase field.  An explicit algebraic derivation of both cases is provided in End Matter. Higher order terms correspond to more complex multipolar structures. 

In the following, we consider specifically solitons in saturable photorefractive crystals operating in the paraelectric phase, where Eq. \eqref{NLSE} reads explicitly as \cite{DelRe1998,DelRe2009,Falsi2024}
\begin{equation} \label{satNLS}
i \, \partial_z \psi + \nabla_{\perp}^2 \psi - \frac{\psi}{(1+|\psi|^2)^2} = 0,
\end{equation}
where $\psi(x,y,z) = E(X,Y,Z)/\sqrt{I_b}$ is the normalized, slowly varying optical field in units of the background intensity $I_b$.
The physical coordinates relate to the dimensionless ones through $(X,Y) = X_0(x,y)$ and $Z = Z_{nl}z$, where the characteristic transverse and longitudinal scales are given by
\[
X_0 = \left(2k_0 n_1 \chi\right)^{-1/2}, \quad Z_{nl} = \chi^{-1}.
\]
Here, $k_0$ is the vacuum wavenumber, $n_1$ the unperturbed refractive index, and the parameter
\[
\chi = \frac{1}{2}k_0 n_1^3 |g_{\mathrm{eff}}| \epsilon_0^2 \left(\epsilon_r(0) - 1\right)^2 E_0^2
\]
includes material and experimental constants: $g_{\mathrm{eff}}$ is the effective quadratic electro-optic coefficient, $\epsilon_0$ the vacuum dielectric constant, $\epsilon_r(0)$ the low-frequency relative dielectric constant, and $E_0$ the applied static bias electric field.  The family of soliton solutions can be numerically identified and mapped as a soliton existence curve in the $(u_0, \Delta X)$ parameter space (see full line in Fig.1d), where $u_0^2=I_0/I_b$ is the so-called intensity ratio, the ratio between the peak beam input intensity and the background illumination, while $\Delta X$ is the input beam FWHM in normalized units (equal for the x and y directions).  The existence curve is the 2+1D extension of previously investigated 1+1D saturated Kerr-like models \cite{Falsi2024}.  In analogy to the 1+1D case, two asymptotic regimes can be identified: a Kerr region for low values of $u_0$, and a saturated region for high values of $u_0$.   

\begin{figure}
    \centering
    \includegraphics[width=1\linewidth]{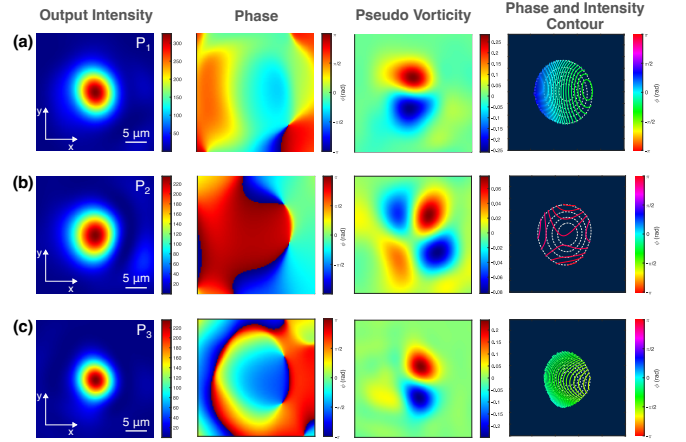}
    \caption{Observed pseudovorticity in photorefractive solitons for input conditions $P_1, P_2, P_3$ defined in Fig. 1d.
Columns show, respectively, output intensity, phase, pseudovorticity, and phase–intensity contours.
P1 (saturated regime) and P3 (Kerr-like regime) both lie on the existence curve and exhibit a dipolar pseudovorticity associated with transverse beam velocity.
P2, outside the existence curve, displays a quadrupolar structure, reflecting the enhanced phase–intensity mismatch in non-self-trapped conditions. Note how the phase and intensity contours (fourth column) retrace the examples illustrated in Fig.1a-c.}
    \label{fig:enter-label}
\end{figure}

\noindent {\it Experiments in photorefractive solitons.} Experiments are performed in a 2.1$^{(x)}$ $\times$ 1.9$^{(y)}$ $\times$ 2.5$^{(z)}$ mm photorefractive KTN:Li sample ($K_{0.964}$Li$_{0.036}$Ta$_{0.60}$Nb$_{0.40}$O$_3$) kept 9K above its $T_C=293$K Curie point, in the paraelectric phase. The zero-cut sample is biased by an $x$-directed static electric field $E_0$, delivered to the sample by electrodes on the $x$ facets. The beam undergoing nonlinear propagation is an x-polarized Gaussian beam from a doubled 30 mW Nd:YAG laser (wavelength $\lambda=2\pi/k_0=532$ nm), focused onto the input $xy$ facet using a spherical lens ($f=75$mm), and travelling along the $z$ axis. Saturation in the nonlinearity is fixed by illuminating the sample from the top (in the $y$ direction) with a uniform background intensity $I_b$. Results refer to steady-state conditions, i.e., those for which the photorefractive nonlinear response no longer depends on time ($t\gg $10 s in our setup). Input and output transverse intensity distributions along the $z$ axis are imaged using a moveable spherical  lens (of focal length $50$mm), onto a CMOS camera.

Experimental input launch conditions leading to beam self-trapping  reported in Fig.1d (magenta circles) are in good agreement with the theoretical prediction of Eq. \eqref{satNLS}. We note that to analyze the effects of saturation on pseudovorticity, we studied 2+1D solitons also in the hereto unexplored region of unsaturated $u_0<1$ Kerr parameter space.  Here, stable 2+1D self-trapping is observed as remnant saturation and spatial nonlocality associated to the nonlinear Raman effect eliminate catastophic self-focusing built into the ideal 2+1D Kerr nonlinearity,
as well known from theory \cite{Zakharov1975,Lemesurier1988,Fibich2015} and experiments in other saturable media \cite{Bjorkholm1974,deAraujo2013}.           

Pseudovorticity of the optical field is experimentally evaluated from the reconstructed complex transverse field distribution. The optical field at the output facet of the crystal is measured using an off-axis interferometer. The interference pattern between the signal beam and a tilted reference beam is recorded on a CMOS camera, allowing for the retrieval of both the amplitude and the phase of the optical field.
The complex field $\psi(x,y)=|\psi(x,y)|e^{i\phi(x,y)}$ is reconstructed by applying a standard phase-retrieval algorithm based on spatial-frequency filtering in the Fourier domain, which isolates the first-order interference term and enables the extraction of the phase distribution ~\cite{Cuche2000}.  
  
Figure 2 reports the output intensity (first column), phase (second column), pseudovorticity (third column), and phase--intensity contours (fourth column) for the three representative input conditions $P_1,P_2,P_3$ (Fig.1d). Compared to the intensity distribution, that remains approximately circular symmetric, and to the phase distribution, that has a complex landscape, pseudovorticity distributions provide clear dipolar/quadrupolar signatures, the result of the misalignment between $\nabla \phi$ and $\nabla \rho$. 
In detail, experimental input conditions that map to points $P_1$ and $P_3$, on the soliton existence curve,  lead to stable saturated ($P_1$) and Kerr-like ($P_3$) 2+1D solitons with a dipolar pseudovorticity.  The dipole reflects the transverse $x$-directed velocity that accompanies soliton formation, a self-bending associated to the nonlinear Raman effect.  
The input condition  $P_2$, lying above the saturated region of the existence curve and not leading to  a self-trapped beam, is accompanied by a more elaborate quadrupolar pseudovorticity pattern. In turn, input conditions below the saturated portion of the existence curve are not accompanied by a  clear pseudovorticity pattern.  Conditions outside the soliton existence curve in the unsaturated Kerr-like regime lead to random-like beam distortion and fragmentation, again with no clear pseudovorticity signature.

\begin{figure}
    \centering\includegraphics[width=1\linewidth]{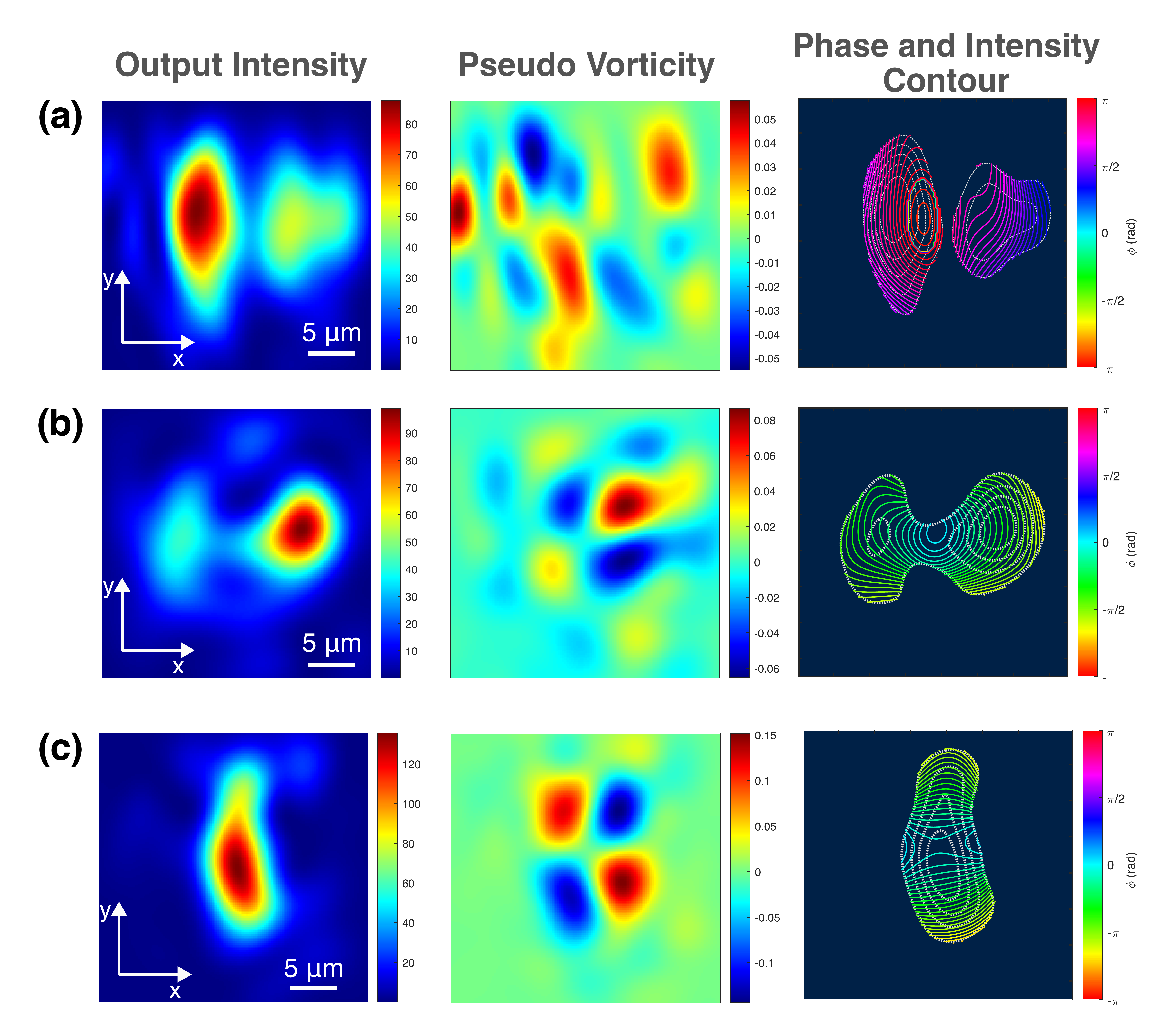}
    \caption{Pseudovorticity in soliton–soliton interactions in the saturated regime ($P_1$ conditions).
(a) Linear propagation: no well-defined pseudovorticity is observed.
(b) Increasing nonlinearity: a quadrupolar pattern progressively emerges.
(c) Soliton fusion: a well-defined quadrupole forms from the superposition of two oppositely oriented dipoles associated with the individual solitons.}
    \label{fig:enter-label}
\end{figure}

In appropriate conditions, solitons in the saturated regime can also  collide and fuse together ~\cite{Xin2019, Xin2021, Xin2022}.  This leads to an interesting scenario for pseudovorticity, as reported in Fig.3.  In detail, two beams with launch conditions $P_1$ are made to collide inside the sample, with an input slanted direction of propagation $\theta_1=-\theta_2\approx 3.9 \times 10^{-3}\ \text{rad}$ relative the $z$ axis. In the absence of nonlinearity (i.e., for $E_0=0$), the output diffraction pattern shown in Fig.3a leads to no defined pseudovorticity distribution. As the nonlinearity builds up, the quadrupolar structure begins to form (see Fig.3b for an intermediate nonlinearity), ultimately leading to a characteristic pseudovorticity quadrupole when fusion occurs in conditions in which each beam  leads to a soliton ($E_0= 1.2 kV/cm$ for this specific case).  This fits in well with the fact that a quadrupole is the natural result of the fusion of two oppositely polarized original single-soliton dipoles (see Fig.3c and description of dynamics of the fusion in End Matter).
   
\begin{figure}
    \centering
    \includegraphics[width=1\linewidth]{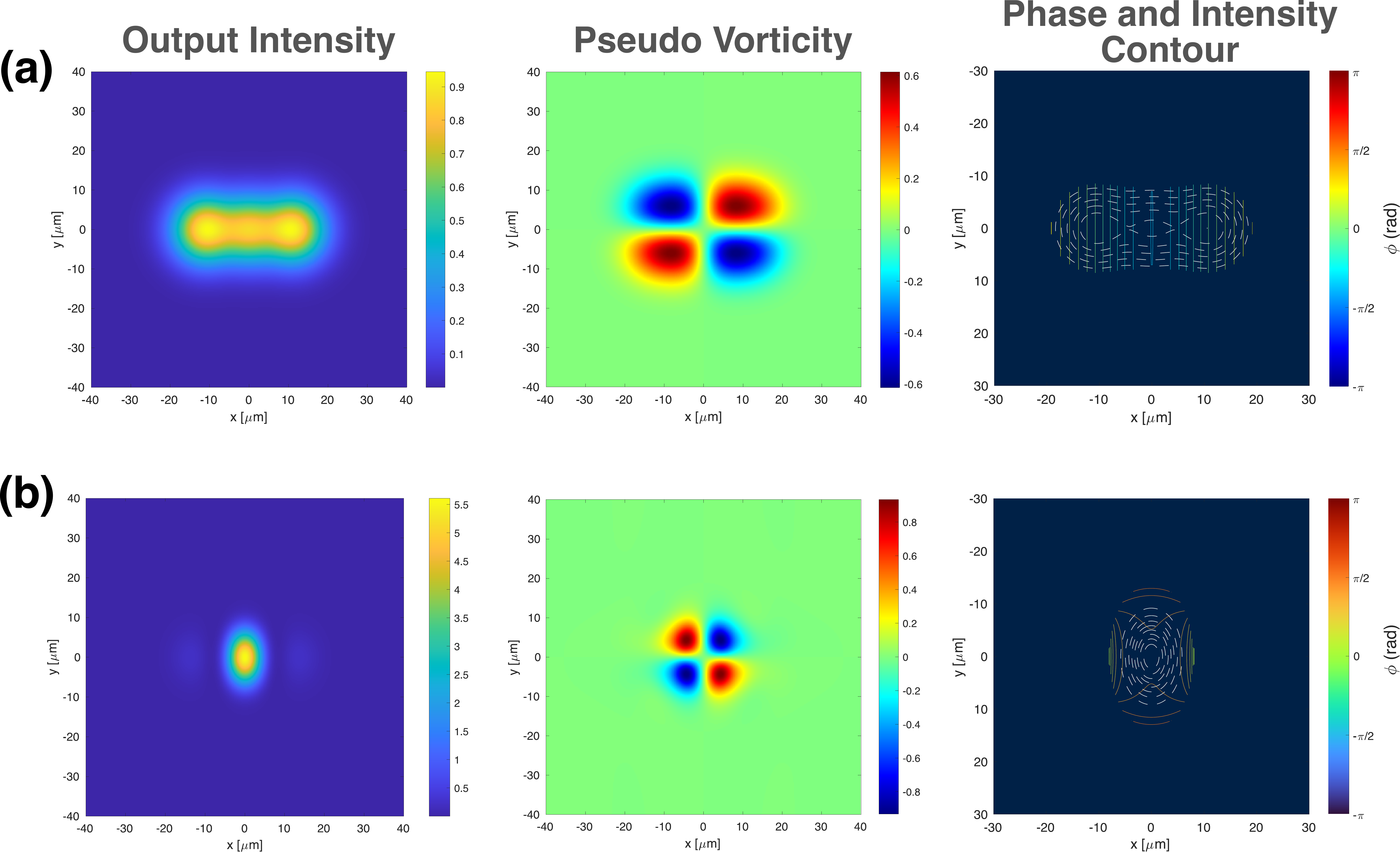}
    \caption{Numerical simulation of soliton fusion.
(a) Linear diffraction regime: two spatially separated dipolar pseudovorticity structures with opposite polarity are found, corresponding to the two independent beams; phase–intensity contours exhibit the expected geometry described in Fig. 1a, with a slight interaction between the two dipoles.
(b) Nonlinear fusion in the saturated regime: a quadrupolar pseudovorticity pattern emerges from the merging of the two dipoles; phase–intensity contours display an elliptical phase profile, consistent with the second-order perturbation (Fig.1c).}
    \label{fig:enter-label}
\end{figure}

Observations can be compared to a beam propagation method simulation of the fusion using the model of Eq.\eqref{satNLS} in the saturated experimental conditions.  As shown in Fig.4, in agreement with experiments, the counterpropagating solitons in the transverse plane fuse, and their pseudovorticity is transformed from a  dipolar structure to a  quadrupolar one. 

{\it Discussion.} To date, optical manipulation techniques make use of OAM and are not based on finite current vorticity, which generally requires specifically tailored beam interference patterns.
In our study, we demonstrate how pseudovorticity patterns naturally accompany spatial soliton formation, allowing, in principle, a hereto unexplored manipulation of objects.
For one, solitons can deliver a torque to confined absorbing particles even though they are topologically neutral: without a singularity, they circumvent the basic limitation of standard particle-rotating techniques based on OAM, i.e., the inability to cause rotation of small particles positioned at the field singularity, where the optical intensity is zero. Moreover, OAM causes rotation as a global effect of the entire field distribution, while soliton pseudovorticity is naturally a dipolar or quadrupolar distribution, the effect being an optical torque field able to generate rotations in different directions depending on the position of the confined absorbing particle.

Note that our present analysis refers solely to the $z$ component of $\mathbf{\omega}_{ps}$, meaning that only rotations in the transverse $xy$ plane are considered, a valid assumption that holds under paraxial conditions. More elaborate phenomena are expected to arise when the full vectorial nature of pseudovorticity is taken into account.

Pseudovorticity also serves to provide a detailed picture of the underlying wave phenomena in nonlinear propagation. In the case of 2+1D photorefractive solitons, the consistent appearance of a pseudovorticity dipole reflects the built-in anisotropic correction in the nonlinearity associated with the nonlinear Raman contribution, even though its direct role in determining self-trapping conditions through Eq.~\eqref{satNLS} is negligible (see Fig.~2a)~\cite{DelRe2005}. This is well illustrated by the formation of a quadrupole for the nonsoliton-supporting input conditions $P_2$ described in Fig.~2b. The quadrupole can be rationalized by considering that, for conditions close to but above the existence curve, i.e., with a $\Delta x$ slightly larger than the soliton-supporting value at the given $u_0$, the beam self-focusing is stronger than diffraction and the beam shrinks from input to output. Superimposing a focusing phase pattern, that is, a parabolic-like phase structure, on a linear chirp associated with the nonlinear Raman effect, a condition can emerge in which, locally, the pattern becomes bell-like, leading to a higher-order perturbation, namely a quadrupole. This is consistent with the fact that an analogous phenomenon does not appear below the existence curve.

More generally, pseudovorticity could play a key dynamical role, i.e., it may be directly connected to local nonlinear wave evolution. For example, pseudovorticity could clarify the mechanism underlying the fission dynamics of rogue waves~\cite{Coppini2026}, where phase–amplitude misalignment could locally induce symmetry breaking and drive the redistribution of energy into multiple localized structures. Indeed, the processes of fusion and fission are closely related by the CPT symmetry of the Schr"odinger equation (see Dynamic description of the fusion in End Matter); consequently, the analysis of pseudovorticity remains valid also in the case of fission.

In summary, experiments in photorefractive materials show that, under conditions leading to 2+1D solitons, strong coupling between phase and amplitude leads to nontrivial dipolar and quadrupolar pseudovorticity patterns. These are associated with the local mismatch of transverse amplitude and phase gradients caused by perturbations that naturally characterize the stable self-trapped state.

{\bf Acknowledgments}
L.F. and E.D.R. acknowledges support the Horizon EIC-Pathfinder Challenges-01 Heisingberg No. 101114978 project. A.V. was funded by Progetti di Ricerca di Interesse Nazionale (PRIN), Project No. 2020X4T57A and 2022WKRYNL. M.O. acknowledges support  from The Simons Foundation (USA) Award 652354 on Wave Turbulence and from INFN (MMNLP and FIELDTUR projects). A.V. acknowledges fruitful discussion with M.J. Cooker. F.C. acknowledges support from PRIN2022
No. 20223T577Z, and from the NSF under grant DMS-2406626.\\

\bibliographystyle{apsrev4-2}
\bibliography{bibliography}

\section{END MATTER}
Soliton solutions of Eq. \eqref{satNLS} can be sought in the form
\begin{equation}\label{bound_state}
\ba{l}
\psi(x,y,z;\lambda)=u(r)e^{i\lambda^2 z}, \ \ r=\sqrt{x^2+y^2}
\ea
\end{equation}
where the family of shape-invariant, radially-symmetric and topologically neutral (flat-phase) profiles $u(r)$ are parametrised by the nonlinear shift $\lambda^2$, and can be numerically found as bound states of the nonlinear eigenvalue problem
\begin{equation}
u''+\frac{u'}{r}-\frac{u}{(1+u^2)^2}=\lambda^2 u,
\end{equation}
subject to boundaries $u(0)=u_0$, $u'(0)=u(\infty)=u'(\infty)=0$. The main features of the soliton family are conveniently summarised by the existence curve in Fig. 1(d), which reports the FWHM of the intensity profile $\rho_s(r) \equiv u^2(r)$ vs. its peak amplitude $u_0=\sqrt{\rho_0}$. This approach can be readily generalised to Eq. \eqref{NLSE} for different nonlinear response $F(|\psi|^2)$.  
 
{\bf Interpretation of the Multipolar structure}\\
Our aim is to propose a mathematical description of the pseudovorticity induced by perturbations of solitons which has a high degree of generality, i.e. it does not rely of the specific model that describes the soliton.
To this end we resort to a local analysis, expanding the transverse dependence in amplitude and phase of the wave-function $\psi(x,y)=\sqrt{\rho(x,y)}e^{i\phi(x,y)}$ around the characteristic point of the system (which corresponds to the peak amplitude in $x=y=0$ in the case of an isolated soliton, whereas, in the study of scattering, it becomes the coalescence point). By performing Taylor expansions, keeping only the minimal number of terms, we obtain
    \begin{equation}
\rho(x,y)\simeq \rho_0-a x^2-b y^2, \ \mbox{\textit{generic localization property}, }
  \end{equation}
with $a,b>0$, whereas for the phase we consider three different cases, say $\phi_{1,2,3}=\phi_{1,2,3}(x,y)$ such that relevant terms are, respectively, odd (linear) or even (quadratic or quartic) in $x,y$:
\begin{eqnarray}
\phi_{1}&\simeq& c_1 x+d_1 y, \hspace{1.8cm} \mbox{\textit{generic phase}}, \nonumber \\
\phi_{2}&\simeq& c_2 x^2+d_2 y^2, \hspace{1.5cm} \mbox{\textit{generic even phase}},\\
\phi_{3}&\simeq& c_3 x^4+d_3 x^2 y^2+e_3 y^4, \mbox{\textit{non generic even phase}}. \nonumber
\end{eqnarray}
	The pseudovorticity field is directed in the out-of-plane \(z\)-direction $\boldsymbol{\omega}_{ps} =\omega_{ps}\,\hat{\mathbf{z}}$. Far from the center, exponential localization of $\rho(x,y)$ is of course important, but not for understanding
	the characteristic pseudovorticity for which the local analysis is sufficient. For the previous modulus and phase configurations, one obtains respectively:
	\begin{equation*}
    \begin{split}
&\omega_{ps_1}(x,y)=2(b c_1 y-a d_1 x), \\
&\omega_{ps_2}(x,y)=4(b c_2-a d_2)xy, \\
&\omega_{ps_3}(x,y)=4xy\left[(2bc_3-ad_3)x^2+(bd_3-2ae_3)y^2\right].
 \end{split}
	\end{equation*}

\textbf{Dipole.}
The nodal line $y=(ad_1/bc_1)x$, i.e. $\omega_{ps_1}(x,y)=0$ partitions the neighborhood of the origin (soliton peak) into two regions, yielding a dipolar configuration with that line as its axis. Figure~1a illustrates, as an example, the specific case of a Gaussian density  $\rho(x,y)$ with $d_1=0$, $c_1\neq 0$.

\textbf{Quadrupole.}
The condition $\omega_{ps_2}(x,y)=0$ along the axes $x=0$ and $y=0$ partitions the neighborhood of the origin into four regions, giving rise to a quadrupolar structure aligned with the coordinate axes. Notably, whether the phase is elliptic ($c_2 d_2>0$) or hyperbolic ($c_2 d_2<0$) does not affect the quadrupolar topology. Figures~1b, c illustrate these two cases, respectively.

For $\omega_{ps_3}(x,y)=0$, two distinct subcases arise.
If $(2 b c_3 - a d_3)(b d_3 - 2 a e_3) > 0$, then $\omega_3 = 0$ only along the coordinate axes $x=0$ and $y=0$. In this case, the neighborhood of the origin is partitioned into four sectors, corresponding again to a quadrupolar structure aligned with the coordinate axes.

Conversely, if $(2 b c_3 - a d_3)(b d_3 - 2 a e_3) < 0$, the condition $\omega_3 = 0$ leads to more intricate nodal patterns, associated with higher-order (octupolar) structures.
Octupolar, or even more complicated patterns could be in principle observable, though, being related to higher-order phase perturbations, are less likely to appear spontaneously in experiments.
   
{\bf Dynamic description of the fusion}\\
We also apply the above preliminaries to describe the pseudovorticity dynamics of the soliton fusion process reported in the paper. 
To this end we consider two colliding solitons with speeds $\pm v$, described by wave-functions $\psi_{\pm}(x,y,z)$, 
simply obtained by applying a Galileian boost to stationary solitons of Eq. \eqref{satNLS} (or, equivalently, Eq. \eqref{NLSE})
\begin{equation} \label{soliton}
\ba{l}
\psi_{\pm}(x,y,z)=e^{\mp i \frac{v}{2}(x\pm \frac{v}{2}z)}u(r_{\pm})e^{i\lambda^2 z}, \\
r_{\pm}=\sqrt{(x\pm v z)^2+y^2}.
\ea\end{equation}

      Let $\psi(x,y,z)$ be the solution of Eq. \eqref{satNLS} describing such a dynamics for distances $z\ll -1$ (assuming coalescence occurs at distance $z=0$)
      since the solitons are exponentially localized, the solution is described by the their sum:
\begin{equation}
\psi(x,y,z)= \psi_{+}(x,y,z)+\psi_{-}(x,y,z), \ \ z \ll z_{fus},
\end{equation}
up to exponentially small corrections. Then $\psi(x,y,z)$ is an even function of $x$ and $y$ for $z\ll -1$, and this property is clearly satisfied $\forall z\in\mathbb{R}$ and, in particular, during the fusion process, that we assume to take place in the origin $(x,y,z)=(0,0,0)$. It follows that the pseudovorticity is an odd function of $x$ and $y$:
\begin{equation}
\ba{l}
\omega_{ps}(-x,y,z)=-\omega_{ps}(x,y,z)=\omega_{ps}(x,-y,z) \\
\Rightarrow \ \omega_{ps}(-x,-y,z)=\omega_{ps}(x,y,z),
\ea
\end{equation}
and can be written as
\begin{equation}
\omega_{ps}(x,y,z)=x\,y\, F(x,y,z),
\end{equation}
where $F(x,y,z)$ is an even function of $x$ and $y$, whose support is localized around the two solitons for $z\ll -1$, and around the origin at fusion.

 \begin{figure}    
    \centering
    \includegraphics[width=1\linewidth]{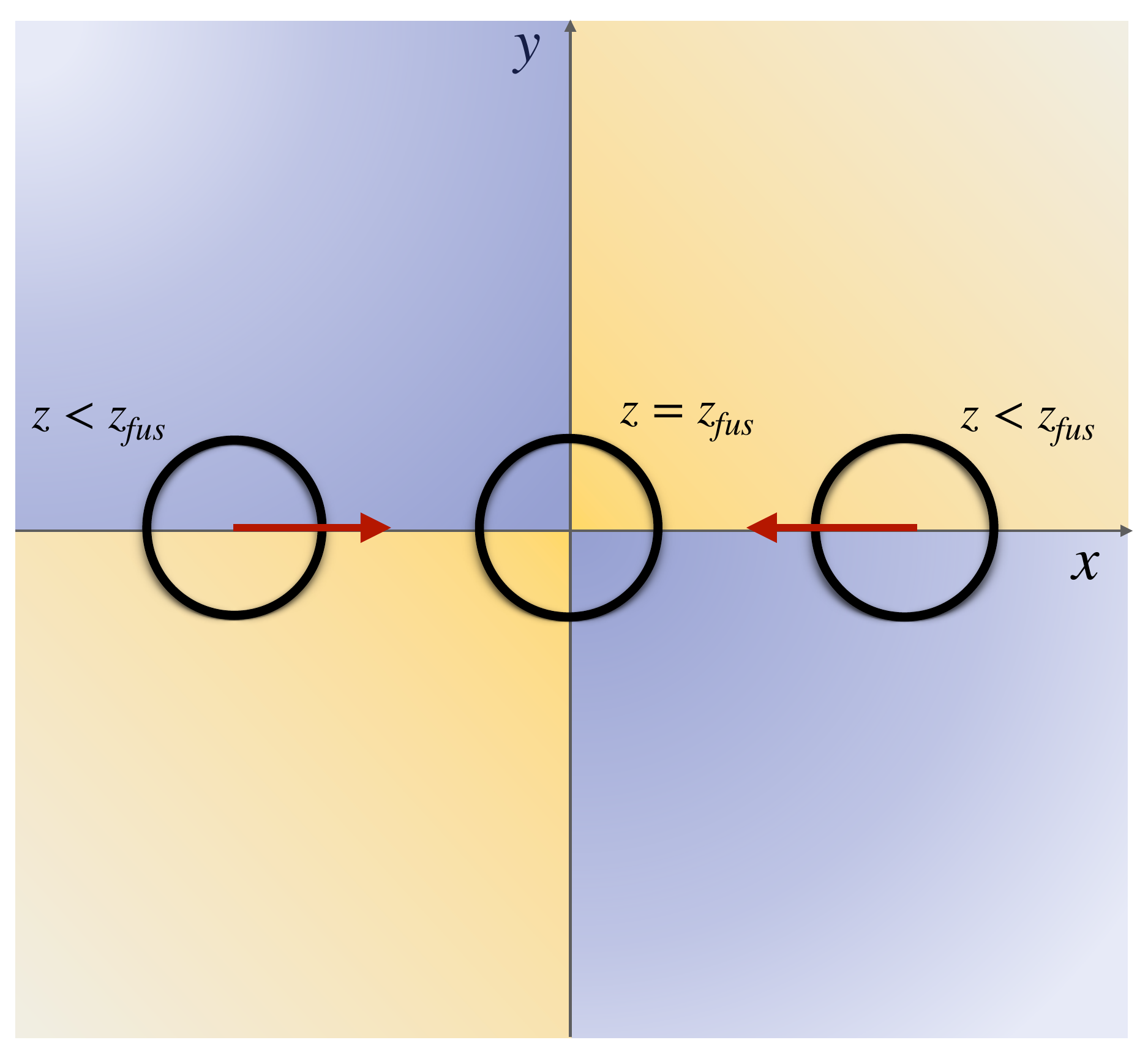}\caption{Schematic illustration of pseudovorticity evolution in the soliton fusion process. Different colors corresponds to pseudovorticity of opposite sign.
    The two circles pointing against each other (red arrows) are solitons far from the coalescence distance, whereas the central circle is the result of the coalescence at $z=z_{fus}$ (see text for more details).}
     \label{fig:fig5}
\end{figure}
For $z\ll -1$, the local phases of the two solitons are linear in $x$ with opposite sign (the overall phase of $\psi(x,y,z)$ depends on $|x|$, as it has to be for parity reasons),  and in agreement with the mechanism of example $\omega_{ps_1}$, the pseudovorticity describes two dipoles traveling towards each other, with nodal $x$ axis and opposite orientations. When the two solitons get close enough and start interacting, the above genericity argument for even wave functions, illustrated in example $\omega_{ps_2}$ and in the first of the two subcases of $\omega_{ps_3}$, suggests that $F(x,y,z)\ne 0$ in a neighborhood  of the origin, so that the pseudovorticity nodal lines are again the $x$ and $y$ axes.

Summarizing, during the $z$ (time) evolution from the remote past to fusion, the pseudovorticity is zero only on the $x$ and $y$ axes, dividing the $(x,y)$ plane into four quadrants of alternating sign (different coloring in Fig.~\ref{fig:fig5}). In the remote past ($z<z_{fus}$) the pseudovorticity support are the two disjoint circles around the centers $(x\pm v z,  y=0)$ of the two solitons, and these two circles are divided by the $x$ axis into two regions of opposite sign, corresponding to two dipoles with opposite orientation (see Fig.~5). At fusion ($z=z_{fus}$), the support, centered at the origin, is divided by the $x$ and $y$ axes into the usual four quadrants; then the pseudovorticity changes sign four times and describes a quadrupole (see Fig.~\ref{fig:fig5}). Although the complicated analytic description of soliton fusion is still missing, the above reasoning, based on elementary geometric considerations, on parity considerations,  and on a local genericity argument, provides a simple justification of the experimental and numerical findings of the paper.

 \end{document}